\documentclass[a4paper,11pt]{article}
\usepackage{pos}
\usepackage{graphicx}
\usepackage{dcolumn}
\usepackage{bm}
\usepackage{hyperref}
\usepackage{orcidlink}
\usepackage{xcolor,amsmath}
\usepackage{flushend}
\usepackage{multirow}
\usepackage{subcaption}
\usepackage{float}
\usepackage{cleveref}
\usepackage{stfloats}
\usepackage{rotating}
\usepackage{comment}

\def \beq{\begin{equation}}
\def \eeq{\end{equation}}
\def \bea{\begin{eqnarray}}
\def \eea{\end{eqnarray}}
\def \bma{\begin{matrix}}
\def \ema{\end{matrix}}

\def \({\left(}
\def \){\right)}
\def \[{\left[}
\def \]{\right]}
\def \nn{\nonumber}

\def \la{\lambda}

\def\tT{{\widetilde{T}}}
\def\tC{{\widetilde{C}}}
\def\tA{{\widetilde{A}}}
\def\tP{{\widetilde{P}}}

\title{Anomalies in Hadronic $B$ Decays}

\author*[a]{Marianne Bouchard}
\author[a]{David London}

\affiliation[a]{Physique des Particules, Universit\'e de Montr\'eal, 1375 Avenue Th\'er\`ese-Lavoie-Roux, Montr\'eal, QC, Canada  H2V 0B3}

\emailAdd{marianne.bouchard.5@umontreal.ca}
\emailAdd{london@lps.umontreal.ca}

\abstract{
The decays $B\to PP$, where the pseudoscalar $P$ is a $\pi$ or $K$, have been studied under the assumption of flavour SU(3) symmetry [SU(3)$_F$]. The global fit shows a 3.6$\sigma$ discrepancy with the Standard Model (SM). Separate fits for $\Delta S = 0$ and $\Delta S = 1$ decays find parameter sets that differ by a factor of 10, suggesting 1000\% SU(3)$_F$ breaking, significantly larger than the $\sim 30\%$ breaking expected in the SM. This study has been extended to include final states with $\eta$ and $\eta'$ mesons. The resulting global fit, once again under the assumption of SU(3)$_F$ symmetry, is worse, with a 4.1$\sigma$ deviation from the SM. When theoretical constraints $|\tC/\tT| = 0.2$ or $\tA = 0$ are imposed, the fits worsen, with the discrepancy approaching 5$\sigma$. These results 
hint at new-physics contributions to these decays.

}

\FullConference{XVI International Conference on Beauty, Charm, Hyperons in Hadronic Interactions (BEACH 2026)\\
7-12 June 2026\\
Firenze, Italy\\}

\begin{document}
\maketitle

\section{Introduction}

The decays $B\to PP$, where $B\in \{ B^+, B^0, B_s^0\}$ and $P\in\{\pi, K\}$, were studied under the assumption of flavour SU(3) symmetry [SU(3)$_F$] in Ref.~\cite{Berthiaume:2023kmp}. Under this symmetry, the decays are related: their amplitudes can all be expressed in terms of a reduced set of unknown parameters. By performing a fit to the available observables, it can be determined how well the Standard Model (SM) explains the data in the SU(3)$_F$ limit. In Ref.~\cite{Berthiaume:2023kmp}, it was found that the individual $\Delta S =0$ and $\Delta S =1 $ fits are good, but the combined fit is very poor, exhibiting a $3.6\sigma$ discrepancy with the SM. The data suggest 1000\% SU(3)$_F$ symmetry breaking, much larger than the value expected within the SM, $\frac{f_K}{f_\pi}-1\simeq20$-$30\%$.

A natural extension to this study is to include the missing pseudoscalar mesons $\eta$ and $\eta'$. This study was conducted in Ref.~\cite{Bhattacharya:2025wcq}. A fit to experimental data was done, again under the assumption of exact SU(3)$_F$ symmetry, and the results reveal a 4.1$\sigma$ discrepancy with the SM. Recently, the study was extended to examine $B\to VV$ decays, where $V\in\{\rho, K^*, \omega, \phi\}$, the excited states of the pseudoscalar mesons. The disagreement with the SM reaches $>7\sigma$ when all the decays are included in the fit \cite{bhattacharya:2026}.

A review of these results was presented at the XVI International Conference on Beauty, Charm, Hyperons in Hadronic Interactions (BEACH 2026), and is reported here.

\section{Observables}\label{sec:obs}

To determine how well the SM can reproduce the experimental data, the predictions for the observables in $B\to PP$ decays are compared with their measured values. These observables include branching ratios (${\cal B}_{CP}$), direct CP asymmetries (${\cal A}_{CP}$) and indirect CP asymmetries (${\cal S}_{CP}$):
\bea
    &{\cal B}_{CP} = F_{\rm PS} \, (|A|^2 + |{\bar A}|^2) ~, &\\
    & {\cal A}_{CP} =\frac{|{\bar A}|^2-|A|^2}{|{\bar A}|^2+|A|^2} ~, & \\
    & {\cal S}_{CP} = \eta_{CP} \times 2\, \text{Im} \, \left(\frac{q}{p}\frac{{\bar A}A^*}{|{\bar A}|^2 + |A|^2}\right) ~, &
    \label{eq:cp_asymmetries}
\eea
where  
$F_{PS}={\sqrt{m_B^2-(m_{P_1}+m_{P_2})^2}\sqrt{m_B^2-(m_{P_1}-m_{P_2})^2}~S}/{(32\pi m^3_B\Gamma_B)}$ is the phase-space factor.
%
%
In these equations, the amplitude for the CP conjugate decay, $\bar A$, can be obtained from the decay amplitude $A$ by changing the sign of the weak phase. $q/p=$ exp$(-2i\phi_M)$, where $\phi_M$ is the phase of $B^0_{q}$-$\bar B^0_{q}$ ($q=d,s$) mixing. 

Under the assumption of SU(3)$_F$ symmetry, the amplitudes can be written in terms of a limited set of unknown parameters common to all decays. There are two choices for this set of unknown parameters. 
The Wigner-Eckart theorem can be used to express the amplitudes in terms of reduced matrix elements (RMEs). Equivalently, the amplitudes can be expressed as functions of topological quark diagrams. These two bases are detailed in the following two sections.

\section{RMEs}\label{sec:RMEs}

With the Wigner-Eckart theorem, we can write the amplitude $\langle PP | H_W | B \rangle$ as the sum of products of an SU(3)$_F$ Clebsch-Gordan coefficient and a reduced matrix element. The initial $B\in\{B^+, B^0, B^0_s\}$ forms a triplet ${\bf 3}$ under SU(3)$_F$. The $B\to PP$ decays are governed by the weak Hamiltonian $H_W$ \cite{Buchalla:1995vs}, and involve the quark-level transitions $\bar{b} \to \bar{u} u \bar{q}$ and ${\bar b} \to ({\bar q} d {\bar d}+{\bar q} s {\bar s})$ ($q = d, s$). The operators in $H_W$ transform as a ${\bf 3^{*}}$, ${\bf 6}$, or ${\bf 15^*}$ of SU(3)$_F$. For the decays with $P\in\{\pi, K\}$, since both pions and kaons are part of the octet ${\bf 8}$ of SU(3)$_F$, they are considered to be identical particles. The final state must therefore be symmetrized:  $(\boldsymbol{8\otimes8})_S$ transforms as $\boldsymbol{1\oplus8\oplus27}$. This leads to five RMEs. 

However, the operators in $H_W$ come in two types, those proportional to $\la_u^{(q)}$ and those proportional to $\la_t^{(q)}$, where 
$\la_u^{(q)} = V_{ub}^* V_{uq}^{}$ and $\la_t^{(q)} = V_{tb}^* V_{tq}^{}$ ($q=d,s)$. Here, the $V_{ij}$ are elements of the Cabibbo-Kobayashi-Maskawa (CKM) matrix. Similarly, each RME can be split into two pieces, one proportional to $\la_u^{(q)}$ and one proportional to $\la_t^{(q)}$:
\bea
& A_1 = \langle {\bf 1} || {\bf 3^*} || {\bf 3} \rangle = \la_u^{(q)} A_1^u - \la_t^{(q)} A_1^t ~,~~
A_8 = \langle {\bf 8} || {\bf 3^*} || {\bf 3} \rangle = \la_u^{(q)} A_8^u - \la_t^{(q)} A_8^t ~, & \nn\\
& R_8 = \langle {\bf 8} || {\bf 6} || {\bf 3} \rangle = \la_u^{(q)} R_8^u -  \la_t^{(q)} R_8^t ~~~~,~~ 
P_8 = \langle {\bf 8} || {\bf 15^*} || {\bf 3} \rangle = \la_u^{(q)} P_8^u - \la_t^{(q)} P_8^t ~, & \nn\\
& P_{27} = \langle {\bf 27} || {\bf 15^*} || {\bf 3} \rangle = \la_u^{(q)} P_{27}^u - \la_t^{(q)} P_{27}^t ~. &
\label{eq:8x8RMEs}
\eea
With this separation, we find that the decay amplitudes can be written in terms of ten RMEs. 

However, they are not all independent. In $H_W$, there are two tree operators, $\lambda_u^{(q)} \sum_{i=1}^{2} c_i Q_i^{(q)}$ and four electroweak penguin (EWP) operators, $\lambda_t^{(q)} \sum_{i=7}^{10} c_i Q_i^{(q)}$. But in the SM, the Wilson coefficients $c_{7,8}$ are only about 5\% as large as $c_{9,10}$ \cite{Buchalla:1995vs}. When $c_{7,8}$ are neglected, $R_8^t$, $P_8^t$, and $P_{27}^t$ are proportional to $R_8^u$, $P_8^u$, and $P_{27}^u$ respectively:
\beq
R_8^t = \frac32 \, \frac{(c_9 - c_{10})}{(c_1 - c_2)} \, R_8^u ~~,~~~~
P_8^t = \frac32 \, \frac{(c_9 + c_{10})}{(c_1 + c_2)} \, P_8^u ~~,~~~~
P_{27}^t = \frac32 \, \frac{(c_9 + c_{10})}{(c_1 + c_2)} \, P_{27}^u ~.
\label{eq:8x8ETR}
\eeq
These are known as EWP-tree relations. In Ref.~\cite{Bhattacharya:2025qye}, it was argued that keeping $c_{7,8}$ modifies these EWP-tree relations by at most $\sim10\%$. For more details on these relations, and for the complete decomposition of the decay amplitudes in terms of RMEs, we refer the reader to Refs.~\cite{Bhattacharya:2025wcq, Bhattacharya:2025qye}.

\section{Diagrams}\label{sec:diagrams}

An alternative approach is to express the amplitudes in terms of topological diagrams \cite{Gronau:1994rj, Gronau:1995hn}. When all the diagrams are kept, the two analyses are equivalent. These diagrams simplify the analysis, since the SU(3)$_F$ Clebsch-Gordan coefficients do not need to be computed. Also, the diagrams represent dynamical processes, allowing us to estimate their relative sizes. For these reasons, diagrams are used in this study.
The complete decomposition of the decay amplitudes in terms of diagrams can be found in Ref.~\cite{Bhattacharya:2025wcq}.

As with RMEs, diagrams can be separated into those proportional to $\la_u^{(q)}$ and those proportional to $\la_t^{(q)}$. The diagrams proportional to $\la_u^{(q)}$ that contribute to the $B\to PP$ decays are $T$ (colour-allowed tree), $C$ (colour-suppressed tree), $A$ (annihilation), and $E$ (exchange). 
The $P$ (penguin) and $PA$ (penguin annihilation) diagrams can be proportional to $\la_u^{(q)}$, $\la_c^{(q)}$, or $\la_t^{(q)}$, depending on the up-type quark in the loop. Using the unitarity of the CKM matrix, the contribution of the $c$ quark can be absorbed, defining two diagrams for each: $P_{uc}$ and $PA_{uc}$ are proportional to $\la_u^{(q)}$, while $P_{tc}$ and $PA_{tc}$ are proportional to $\la_t^{(q)}$.
The electroweak penguin diagrams are also proportional to $\la_t^{(q)}$. There is one electroweak penguin corresponding to each diagram proportional to $\la_u^{(q)}$ \cite{Bhattacharya:2025wcq, Gronau:1998fn}: $P_{EW}$, $P_{EW}^{C}$, $P_{EW}^{A}$, $P_{EW}^{E}$, $P_{EW}^{P_u}$, and $P_{EW}^{PA_u}$.

With six diagrams proportional to $\la_u^{(q)}$ and eight proportional to $\la_t^{(q)}$, we have many more diagrams than RMEs. This is because not all diagrams are independent; only certain combinations of diagrams appear in the amplitudes. In fact, there are only five independent combinations of diagrams proportional to $\la_u^{(q)}$, defined as $\tT$, $\tC$, $\tA$, $\tP_{uc}$, and $\widetilde{PA}_{uc}$ (the exchange diagram $E$ has been absorbed into the other contributions). Similarly, there are five diagrams proportional to $\la_t^{(q)}$: $\tP_{tc}$, $\widetilde{PA}_{tc}$, $\widetilde{P_{EW}}$, $\widetilde{P_{EW}^C}$, and $\widetilde{P_{EW}^A}$. We stress that the two bases are equivalent; the RMEs can be written as functions of these effective diagrams \cite{Bhattacharya:2025wcq}. 

The EWP-tree relations can also be written in terms of diagrams. These relations allow us to write $\widetilde{P_{EW}}$, $\widetilde{P_{EW}^C}$, and $\widetilde{P_{EW}^A}$ as functions of $\tT$, $\tC$, and $\tA$ \cite{Bhattacharya:2025wcq, Bhattacharya:2025qye}. This reduces the number of parameters to fit from 10 complex amplitudes to seven. The magnitudes and relative strong phases of these seven effective diagrams will be the unknown parameters for the fits.

\section{Fits to the Data}\label{sec:fit}

A first fit is done using final states with only $\{\pi, K\}$. There are 30 observables available for these decays, 15 for $\Delta S = 0$ and 15 for $\Delta S =1$.
The complete list of observables used in the fits can be found in Ref.~\cite{Bhattacharya:2025wcq}.
All the fits were done with the {\it{Minuit}} package \cite{James:1975dr} to find the values of diagrams that minimize the $\chi^2$ function. 
As mentioned above, the amplitudes are described in terms of seven effective diagrams: $\tT$, $\tC$, $\tA$, $\tP_{uc}$, $\widetilde{PA}_{uc}$, $\tP_{tc}$, and $\widetilde{PA}_{tc}$. This corresponds to 13 unknown parameters (seven magnitudes and six relative strong phases). 

We can start by studying separately the $\Delta S =0$ and $\Delta S =1$ decays. 
With 15 observables for each sector and 13 unknown parameters, fits can be done.
In the $\Delta S=0$ sector, the fit is excellent: $\chi^2_{\text{min}}/\rm d.o.f.=1.1/2$, which corresponds to a $p$-value of 58\%. The fit for $\Delta S=1$ decays is slightly worse, but still perfectly acceptable, with $\chi^2_{\text{min}}/\rm d.o.f.=1.6/2$, or a $p$-value of 45\%. A combined $\Delta S = 0$ and $\Delta S = 1$ fit can also be done when assuming perfect SU(3)$_F$ symmetry. With this assumption, the diagrams for the two sectors are taken to be the same. The fit is now quite poor, with $\chi^2_{\text{min}}/\rm d.o.f.=43/17$. This corresponds to a $p$-value of $4.5\times10^{-4}$, or a 3.6$\sigma$ disagreement with the SM. 

What is the reason for this poor fit? This can be understood by comparing the results of the individual $\Delta S =0$ and $\Delta S =1$ fits. In Table~\ref{tab:fit}, we show the best-fit values of the magnitudes of the diagrams in the two sectors. We see that they differ by a factor $\sim10$. This can be seen particularly in the magnitudes of $\tT$, $\tC$, and $\tP_{uc}$:
\beq
|\tT'/\tT| = 13.1 \pm 2.1 ~,~~~~ |\tC'/\tC| = 8.9 \pm 1.3 ~,~~~~|\tP_{uc}'/\tP_{uc}| =  22\pm28 ~,
\eeq
where the diagrams in $\Delta S=1$ decays are denoted by primes. This corresponds to a 1000\% breaking of SU(3)$_F$ symmetry, far larger than the $\sim30\%$ breaking expected in the SM.

\begin{table}[ht]
\begin{center}
\begin{tabular}{|c|c|c|c|} \hline 
\multicolumn{1}{|c|}{Fit} & $\chi^2_{\text{min}}/\rm d.o.f. = 1.1/2$ & $\chi^2_{\text{min}}/\rm d.o.f. = 1.6/2$ & $\chi^2_{\text{min}}/\rm d.o.f. = 43/17$\\ \cline{2-4}
\multicolumn{1}{|c|}{Parameters} & $\Delta S = 0$ & $\Delta S = 1$ & Combined fit \\ \hline \hline
        $|\widetilde T|$   & $4.2 \pm0.5$   & $ 55 \pm 6$     & $5.5\pm 0.6$\\
        $|\widetilde C|$   & $6.3\pm0.5$  & $ 56 \pm 7$     & $4.7\pm0.4$ \\
        $|\widetilde P_{uc}|$   & $2.7\pm3.4$  & $59 \pm 8$  & $1.4\pm0.6$  \\
        $|\widetilde A|$   & $0\pm8$  & $33\pm 34$  & $3.3\pm0.7$ \\
        $|\widetilde{PA}_{uc}|$& $0.6 \pm0.7$   & $ 3.7 \pm 1.4$     & $0.69\pm0.09$ \\
        $|\widetilde P_{tc}|$   & $0.8\pm1.0$  & $0.73 \pm 0.10$   & $1.184\pm0.008$\\ 
        $|\widetilde{PA}_{tc}|$   & $0.2\pm0.5$  & $0.18 \pm 0.13$   & $0.216\pm0.015$ \\ \hline
\end{tabular}
\end{center}
\caption{Best-fit values for the magnitudes of diagrams in the decays $B\to PP$, $P\in\{\pi,K\}$. Values are given in units of keV.\label{tab:fit}}
\end{table}

\section{Theoretical Input}\label{sec:theory_input}

In the above analysis, the only theoretical input has been to neglect $c_{7,8}$. But their inclusion modifies the EWP-tree relations by only $\sim 10\%$ \cite{Bhattacharya:2025qye}, so the results are close to rigorous, group-theoretically. Even so, it is interesting to examine how the results change when other theoretical input is added.

One advantage of using topological diagrams for the fits is that, even if they are not Feynman diagrams, they still represent dynamical ways in which a decay can take place. This allows us to estimate the relative sizes of these diagrams. For instance, simply by counting colours, we expect the magnitude of $C$ to be $\sim1/3$ that of $T$. This is supported by QCD factorization, which finds $|C/T|\simeq 0.2$ \cite{Beneke:2001ev, Bell:2007tv, Bell:2009nk, Beneke:2009ek, Bell:2015koa}. Since we also expect $E\ll T,C$, the following relation should also hold: $|\tC/\tT|\simeq 0.2$. However, from Table~\ref{tab:fit}, we see that the preferred value in the combined fit is significantly larger: $|\tC/\tT|=0.85\pm0.12$. If the combined fit is repeated with the constraint $|\tC/\tT|=0.2$ imposed, the fit worsens quite a bit, reaching a 4.9$\sigma$ disagreement with the SM.

Another common expectation is that the $A$, $E$, and $PA$ diagrams should be considerably suppressed compared to $T$, $C$, or $P$. Since they involve the interaction of the $\bar b$ quark with the spectator quark, this suppression should be of order $f_B/m_B\simeq5\%$ \cite{Gronau:1994rj}. Because of this, they are often neglected in fits. 
However, in Table~\ref{tab:fit}, we see that, for the combined fit, $|\tA|\sim|\tT|,|\tC|$, which disagrees with this prediction.
If the combined fit is redone while imposing $\tA=0$, the discrepancy with the SM reaches 5.2$\sigma$, an important increase.

\section{\boldmath $\eta$ and $\eta'$}\label{sec:eta}

It is natural to extend the above analysis, which focuses on decays with only $\pi$ and $K$ in the final state, to include the two missing pseudoscalar mesons, $\eta$ and $\eta'$. 
The final-state mesons are now members of an SU(3)$_F$ octet ${\bf 8}$, $\{\pi^\pm, \pi^0, K^\pm, K^0, {\bar K}^0, \eta_8 \}$, and a singlet ${\bf 1}$, $\eta_1$. There are now three types of final $PP$ states, which transform as $(\boldsymbol{8\otimes 8})_S$, $\boldsymbol{8\otimes 1}$, and $\boldsymbol{1\otimes 1}$. 
As above, the $(\boldsymbol{8\otimes 8})_S$ state has seven effective diagrams.
For the $\boldsymbol{8\otimes 1}$ final state, once EWP-tree relations are taken into account, four additional effective diagrams are required.
With the $\boldsymbol{1\otimes 1}$ final states, there are two additional effective diagrams.
For more details, we refer the reader to Ref.~\cite{Bhattacharya:2025wcq}.

The physical particles $\eta$ and $\eta'$ are admixtures of the octet and singlet members $\eta_8$ and $\eta_1$:
\begin{equation}
    \eta = \eta_8 \cos \theta_\eta - \eta_1\sin \theta_\eta ~~,~~~~
    \eta' = \eta_8\sin\theta_\eta + \eta_1 \cos\theta_\eta ~,
\label{thetaetadef}
\end{equation}
with
\begin{equation}
    \eta_8 \equiv \frac{(2s{\bar s}-u{\bar u} -d{\bar d})}{\sqrt{6}}~~, ~~~~ \eta_1 \equiv \frac{(u{\bar u} + d{\bar d} + s{\bar s})}{\sqrt{3}} ~.
\end{equation}
The value $\theta_\eta = {\rm arcsin}(1/3) \approx 19.5^\circ$ is commonly used and has been adopted in this study. Note, however, that there exist other ways to parametrize the $\eta$ and $\eta'$ mesons \cite{Bolognani:2024zno}.

Decays with one $\pi$ or $K$ and one $\eta$ or $\eta'$ in the final state contribute to $(\boldsymbol{8\otimes 8})_S$ and $\boldsymbol{8\otimes 1}$. Decays with two final-state $\eta$ or $\eta'$ mesons are part of all three sectors, $(\boldsymbol{8\otimes 8})_S$, $\boldsymbol{8\otimes 1}$, and $\boldsymbol{1\otimes 1}$.

We can include decays with no $\eta^{(')}$, one $\eta^{(')}$, or two $\eta^{(')}$s in the final state. 
In total, there are 13 effective diagrams, which corresponds to 25 real parameters for the fit (13 magnitudes and 12 relative strong phases). Individual $\Delta S = 0$ and $\Delta S =1$ fits cannot be performed since these new decays add more unknown parameters than measurements. However, the combined fit can be done, and the results show $\chi^2_{\text{min}}/\text{d.o.f}= 61/24$, which corresponds to a $p$-value of $4.9\times10^{-5}$, or a $4.1\sigma$ discrepancy with the SU(3)$_F$ limit of the SM.

We see that, by including more decays, the discrepancy with the SM increases, reaching over 4$\sigma$ when all the pseudoscalar mesons are included.

\section{Conclusion}\label{sec:conclusion}

The $B\to PP$ decays, $P\in\{\pi,K\}$, are studied under the assumption of flavour SU(3) symmetry, and a global fit to the latest experimental data is carried out. The individual $\Delta S =0$ and $\Delta S=1$ fits are good. The two sectors can be combined in a single fit in the limit of exact SU(3)$_F$ symmetry, and the results show a 3.6$\sigma$ discrepancy with the SM. This is due to the difference between the best-fit values of the magnitudes of diagrams in the two individual fits, which suggests 1000\% SU(3)$_F$ breaking, far larger than the $\sim30\%$ breaking predicted by the SM. When constraints on the diagrams are added ($|\tC/\tT|=0.2$, $\tA=0$), the fit worsens considerably, reaching  $\sim5\sigma$. 

With only pions and kaons, the final state transforms exclusively as $(\boldsymbol{8\otimes 8})_S$ under SU(3)$_F$. However, the complete set of pseudoscalar mesons includes an octet and a singlet. The physical states $\eta$ and $\eta'$ are admixtures of the octet $\eta_8$ and singlet $\eta_1$. These missing pseudoscalar mesons can be added to the fit with the inclusion of extra diagrams for the new final states $\boldsymbol{8\otimes 1}$ and $\boldsymbol{1\otimes 1}$. When $\eta$ and $\eta'$ are added, the discrepancy with the SU(3)$_F$ limit of the SM increases to 4.1$\sigma$. 

The study can be extended to $B\to VV$ decays, where $V$ is a vector meson. The vector mesons have spin 1, so the final state can have spin $S_{tot}=0,1,2$. With this change, the decay amplitudes are expressed in terms of one amplitude for each polarization, so three amplitudes in total. Because there is one diagram per polarization, there are three times as many parameters to fit. More parameters also means more observables for the fit. These new observables include polarization fractions, polarization-dependent direct CP asymmetries, phases differences and many more. This study has been done in Ref.~\cite{bhattacharya:2026}. While there is not enough data to do separate $\Delta S = 0$ and $\Delta S = 1$ fits, a combined fit can be done.
When all the vector mesons are included ($V\in\{\rho,K^*, \omega, \phi\}$), the fit results show a $>7\sigma$ deviation from the SM predictions.

While it seems unlikely that a $\sim30\%$ SU(3)$_F$ breaking could explain all of these anomalies, this possibility has to be checked. This is work in progress.

\bigskip\bigskip
\noindent
{\bf Acknowledgements:} This work was financially supported by NSERC of Canada  (M.B., D.L.), and by FRQNT, Scholarship No.\ 363240 (M.B.).

\bibliographystyle{JHEP}
\bibliography{proceeding}

\end{document}